\documentclass[longauth,traditabstract,letter]{aa}  
\usepackage{graphicx}
\usepackage{txfonts}
\begin{document}

   \title{Aperture corrections for disk galaxy properties derived from the CALIFA survey}

   \subtitle{Balmer emission lines in spiral galaxies}

   \author{J. Iglesias-P\'{a}ramo
          \inst{1,2}
          \and
          J.M. V\'{\i}lchez\inst{1}
          \and
          L. Galbany\inst{3}
          \and
          S.F. S\'{a}nchez\inst{1,2}
          \and
          F.F. Rosales-Ortega\inst{4}
          \and
          D. Mast\inst{2}
          \and
          R. Garc\'{\i}a-Benito\inst{1}
          \and
          B. Husemann\inst{5}
          \and
          J.A.L. Aguerri\inst{6,7}
          \and
          J. Alves\inst{8}
          \and
          S. Bekerait\'{e}\inst{5}
          \and
          J. Bland-Hawthorn\inst{9}
          \and
          C. Catal\'{a}n-Torrecilla\inst{10}
          \and
          A.L. de Amorim\inst{1,11}
          \and
          A. de Lorenzo-C\'{a}ceres\inst{6,7}
          \and
          S. Ellis\inst{12}
          \and
          J. Falc\'{o}n-Barroso\inst{6,7}
          \and
          H. Flores\inst{13}
          \and
          E. Florido\inst{14,15}
          \and
          A. Gallazzi\inst{16,17}
          \and
          J.M. Gomes\inst{18}
          \and
          R.M. Gonz\'{a}lez Delgado\inst{1}
          \and
          T. Haines\inst{19}
          \and
          J.D. Hern\'{a}ndez-Fern\'{a}ndez\inst{20}
          \and
          C. Kehrig\inst{1}
          \and
          A.R. L\'{o}pez-S\'{a}nchez\inst{12,21}
          \and
          M. Lyubenova\inst{22}
          \and
          R.A. Marino\inst{23}
          \and
          M. Moll\'{a}\inst{24}
          \and
          A. Monreal-Ibero\inst{1}
          \and
          A. Mour\~{a}o\inst{3}
          \and
          P. Papaderos\inst{18}
          \and
          P. S\'{a}nchez-Bl\'{a}zquez\inst{25}
          \and
          K. Spekkens\inst{26}
          \and
          V. Stanishev\inst{3}
          \and 
          G. van de Ven\inst{22}
          \and
          C.J. Walcher\inst{5}
          \and
          L. Wisotzki\inst{5}
          \and
          S. Zibetti\inst{17}
          \and
          B. Ziegler\inst{8}
          }

   \institute{Instituto de Astrof\'{\i}sica de Andaluc\'{\i}a - CSIC, 18008 Granada, Spain
         \and
             Centro Astron\'{o}mico Hispano Alem\'{a}n, 04004 Almer\'{\i}a, Spain
         \and
CENTRA – Centro Multidisciplinar de Astrof\'{\i}sica, Instituto
Superior T\'{e}cnico, Av. Rovisco Pais 1, 1049-001 Lisbon, Portugal
         \and
Instituto Nacional de Astrof\'{\i}sica, \'{O}ptica y Electr\'{o}nica, Luis E. Erro 1, 72840 Tonantzintla, Puebla, M\'{e}jico
         \and
Leibniz-Institut f\"{u}r Astrophysik Potsdam (AIP), An der Sternwarte
16, 14482 Potsdam, Germany
         \and
Instituto de Astrof\'{\i}sica de Canarias, V\'{\i}a L\'{a}ctea s/n, La Laguna,
Tenerife, Spain
         \and
Departamento de Astrof\'{\i}sica, Universidad de La Laguna, 38205
La Laguna, Tenerife, Spain
        \and
University of Vienna, Department of Astrophysics,
T\"{u}rkenschanzstr. 17, 1180 Vienna, Austria
         \and
Sydney Institute for Astronomy, School of Physics A28, University
of Sydney, NSW 2006, Australia
         \and
Departamento de Astrof\'{\i}sica y CC. de la Atm\'{o}sfera, Universidad
Complutense de Madrid, 28040 Madrid, Spain
         \and
Departamento de F\'{\i}sica, Universidade Federal de Santa Catarina,
PO Box 476, 88040-900 Florianópolis, SC, Brazil
         \and
Australian Astronomical Observatory, 105 Delhi Rd., North Ryde, NSW 1670, Australia
        \and
GEPI, Observatoire de Paris, CNRS-UMR8111, Univ. Paris-Diderot, 5 place Janssen, 92195 Meudon, France
         \and
Departamento de F\'{\i}sica Te\'{o}rica y del Cosmos, University of
Granada, Facultad de Ciencias (Edificio Mecenas), 18071 Granada,
Spain
         \and
Instituto Carlos I de F\'{\i}sica Te\'{o}rica y Computaci\'{o}n, 18071 Granada,
Spain
         \and
Dark Cosmology Centre, Niels Bohr Institute, University of
Copenhagen, Juliane Mariesvej 30, 2100 Copenhagen, Denmark
        \and
INAF – Osservatorio Astrofisico di Arcetri – Largo Enrico Fermi,
5 – 50125 Firenze, Italy
         \and
Centro de Astrof\'{\i}sica and Faculdade de Ci\^{e}ncias, Universidade do
Porto, Rua das Estrelas, 4150-762 Porto, Portugal
         \and
Department of Physics and Astronomy University of Missouri –
Kansas City, Kansas City, MO 64110, USA
        \and
Departamento de Astronomia do Instituto de Astronomia, Geof\'{\i}sica e Ci\^{e}ncias Atmosf\'{e}ricas da USP, 
Rua do Mat\~{a}o 1226, Cidade Universit\'{a}ria, 05508-090, S\~{a}o Paulo, Brasil
        \and 
Department of Physics and Astronomy, Macquarie University, NSW
2109, Australia
        \and
Max-Planck-Institut f\"{u}r Astronomie, K\"{o}nigstuhl 17, 69117 Heidelberg, Germany
        \and
CEI Campus Moncloa, UCM-UPM, Departamento de Astrof\'{\i}sica
y CC. de la Atm\'{o}sfera, Facultad de CC. F\'{\i}sicas, Universidad
Complutense de Madrid, Avda. Complutense s/n, 28040 Madrid,
Spain
        \and
Departamento de Investigaci\'{o}n B\'{a}sica, CIEMAT, Avda. Complutense 22, 28040 Madrid, Spain 
        \and
Departamento de F\'{\i}sica Te\'{o}rica, Universidad Aut\'{o}noma de Madrid,
28049 Madrid, Spain
        \and
Department of Physics, Royal Military College of Canada, PO Box
17000, Station Forces, Kingston, Ontario, K7K 7B4, Canada
             }

   \date{Received September 15, 1996; accepted March 16, 1997}

 
  \abstract
{This work investigates the effect of the aperture size on derived galaxy properties for which we have spatially-resolved optical spectra.
We focus on some indicators of star formation activity and dust attenuation for spiral galaxies that have been widely used in previous work on galaxy evolution.
We have used 104 spiral galaxies from the CALIFA survey for which 2D spectroscopy with complete spatial coverage is available. 
From the 3D cubes we have derived growth curves of the most conspicuous Balmer emission lines (H$\alpha$, H$\beta$) for circular apertures of different radii centered at the galaxy's nucleus after removing the underlying stellar continuum.
We find that the H$\alpha$ flux ($f$(H$\alpha$)) growth curve follows a well defined sequence with aperture radius showing low dispersion around the median value. From this analysis, we derive aperture corrections for galaxies in different magnitude and redshift intervals.
Once stellar absorption is properly accounted for, the $f$(H$\alpha$)/$f$(H$\beta$) ratio growth curve shows a smooth decline, pointing towards the absence of differential dust attenuation
as a function of radius. Aperture corrections as a function of the radius are provided in the interval [0.3,2.5]$R_{50}$. 
Finally, the H$\alpha$ equivalent width ($EW$(H$\alpha$)) growth curve increases with the size of the aperture and shows a very large dispersion for small apertures.
This large dispersion prevents the use of reliable aperture corrections for this quantity.
In addition, this result suggests that separating star-forming and quiescent galaxies based on observed $EW$(H$\alpha$) through small apertures is likely to result 
in low $EW$(H$\alpha$) star-forming galaxies begin classified as quiescent.
}
   \keywords{techniques: spectroscopic - galaxies: evolution - surveys - galaxies: ISM}

   \maketitle
%

\section{Introduction}

The advent of large area surveys has produced an enormous advance in our knowledge of galaxy formation and evolution in recent decades.
In particular, those consisting of imaging and single fiber spectroscopy (e.g. 2dFGRS, Folkes et al. 1999; SDSS, York et al. 2000; GAMA, Driver et al. 2011) have provided invaluable information on redshift as well as other galactic properties such as star formation rates (SFR) or metallicities, via the intensities of the most conspicuous emission lines.
However, aperture effects are always present in such studies due to the limited (and variable with redshift) coverage of the individual galaxies. 
Consequently, a fraction of the total flux at all wavelengths is lost and, to date, no recipe to correct for this effect has proved satisfactory, although several studies of aperture effects have been
carried out (e.g. Brinchmann et al. 2004; Ellis et al. 2005; Kewley et al. 2005; 
Salim et al. 2007; Kronberger et al. 2008; Gerssen et al. 2012; Zahid et al. 2013; Hopkins et al. 2013).

In practice, there are two main issues: one is the size of the projection on the sky of the aperture 
relative to the physical dimensions of the galaxy. 
The second one, which affects mainly single fiber (or very small aperture) spectrographs (e.g. 2'' for 2dFGRS and GAMA; 3'' for SDSS), 
is the precise position of the aperture relative to the galaxy center.
This aperture effect has implications for spectral stacking studies, since the fraction of a galaxy covered by a fixed aperture varies with redshift.
In this sense, the use of a realistic aperture correction is a {\em sine qua non} condition in order
to carry out the analysis of aperture spectra of galaxies at different epochs, to derive their fundamental properties and to discuss their evolution.

Gerssen et al. (2012) have recently carried out a study of a reduced sample of SDSS star-forming galaxies using the VIMOS integral field spectrograph (Le F\`{e}vre et al. 2003) to map (with full spatial coverage) some properties related to star formation diagnostics. 
These authors found a large dispersion when comparing their results to color-based SDSS extrapolations. 
This suggests that full spatial coverage is needed to produce proper corrections for emission lines intensities, 
and large samples of galaxies covering all the morphological types are required.

The {\em Calar Alto Legacy Integral Field Area} Survey (CA\-LI\-FA; S\'{a}nchez et al. 2012a) is observing a statistically well-defined sample of 600 galaxies in the local Universe with the Potsdam Multi Aperture Spectrograph (PMAS; Roth et al. 2005) in the PPAK mode. 
Galaxies have been selected from the SDSS-DR7 catalog (Abazajian et al. 2009) to fulfill the following conditions: (1) they fall in the redshift range $0.005 < z < 0.03$; and (2) their isophotal diameters in the $r'$ band ({\em isoA\_r} in SDSS-DR7) are in the range $45'' < isoA\_r < 80''$. 
The combination of these three criteria results in a sample of galaxies that fall completely within the PPAK field-of-view with all the relevant emission lines covered with a single spectral setup.
The CALIFA galaxies completely sample the color-magnitude diagram 
down to $M_{r} = -19$~mag (see S\'{a}nchez et al. 2012a).
In summary, the data provided by CALIFA are well suited to (a) testing the aperture effects discussed above, thanks to complete spectral and spatial coverage of the sample galaxies, and (b) providing aperture correction factors as a function of aperture size for quantities that vary smoothly with the latter.
In our analysis, we adopt the following cosmological parameters: $H_{0}=72$~km~s$^{-1}$~Mpc$^{-1}$, $\Omega_{M}=0.27$ and $\Omega_{\lambda}=0.73$.


\section{Aperture corrections}

The basic tools for this process are the CALIFA reduced cubes from the DR1 sample 
(Husemann et al. 2013).
Our starting point is a larger sample containing 258 galaxies including those of DR1 and some others not yet released to the community. The survey sample comprises 
24.4\% E-S0 and 75.6\% Sa-Sm galaxies. 
We remove those galaxies suspected of being AGN according to several criteria based on optical spectra and radio data following the criteria of Best et al. (2012), Kewley et al. (2001, 2006) and Cid Fernandes et al. (2011).
We also removed galaxies with morphological types earlier than S0a since we want to focus only in spiral galaxies.
Given that we want to keep only galaxies completely observed within the PPAK field-of-view, we rejected all the galaxies for which $2.5 R_{50} \geq 36''$, 
where $R_{50}$ is the radius containing 50\% of the Petrosian flux in the $r'$ band ($petroR50\_r$ in SDSS-DR7). 
After testing several options like half light radius from circular and elliptical and circular growth curve analysis we decided to use $petroR50\_r$ as our scale as it results in (a) the largest sample of galaxies fulfilling our previous requirements, and (b) the least dispersed profiles for the growth curves to be discussed later.
After these considerations, we ended up with a sample of 104 galaxies with morphological types from Sa to Sm completely covered by the PPAK field-of-view.
The median values ($1\sigma$ confidence interval) of the stellar mass, SFR (from dust corrected H$\alpha$ luminosities), $EW$(H$\alpha$) and $f$(H$\alpha$)/$f$(H$\beta$) distributions of the sample are $\log M^{*}/M_{\odot} = 10.25$ ($[9.65,10.85]$), $\log SFR/(M_{\odot}~yr^{-1}) = 0.45$ ($[-0.06,0.79]$), $EW$(H$\alpha$)/\AA\ $= 16.91$ ($[6.39,28.56]$) and $f$(H$\alpha$)/$f$(H$\beta$) = 4.75 ($[4.00,5.89]$).
For each galaxy, we produced stacked spectra by adding the emission from the spaxels within circular apertures centered at the galaxy's nucleus 
and with radii increasing from 1.5'' by steps of 1.5'' up to a maximum radius completely covering the hexagonal PPAK aperture.
Then we subtracted the stellar continuum to each of our stacked spectra by using the FIT3D code (S\'{a}nchez et al. 2011), resulting in a set of spectra of the ionized gas emission.
Finally, the fluxes of the emission lines within each aperture were extracted from two independent gaussian fits with the IDL-based routine MPFITEXPR (Markwardt 2009), one for H$\beta$ and another one for the triplet H$\alpha$ $+$ [N{\sc ii}]$\lambda$6548,6583\AA.
A more detailed description on this point will be included in a forthcoming paper.

   \begin{figure}[!t]
   \centering
\includegraphics[width=9cm]{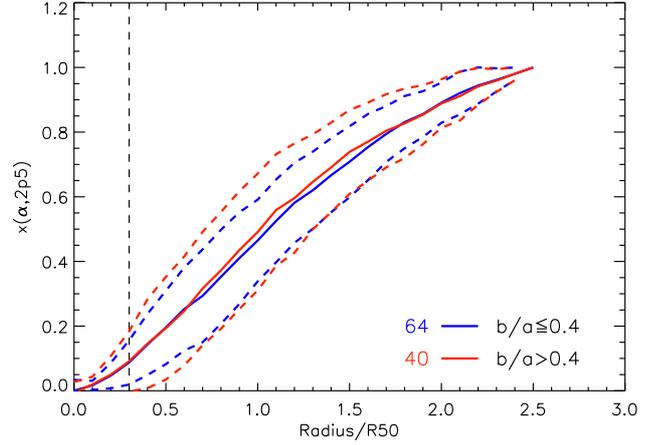}
   \caption{Growth curve of $f$(H$\alpha$) normalized to $f$(H$\alpha$) at $2.5 R_{50}$, $x(\alpha,2p5)$, as a function of the radius of the aperture. Bold red (blue) lines correspond to the median values for galaxies with $b/a > 0.4$ ($b/a < 0.4$). Dashed red and blue lines correspond to $1\sigma$ deviations from the median curve. Numbers indicate the size of each subsample. The vertical dashed line corresponds to $0.3R_{50}$, which on average corresponds to the FWHM of the CALIFA psf for our sample.}
              \label{ha}%
    \end{figure}

\section{Results}

\begin{figure}[!b]
   \centering
\includegraphics[width=9cm]{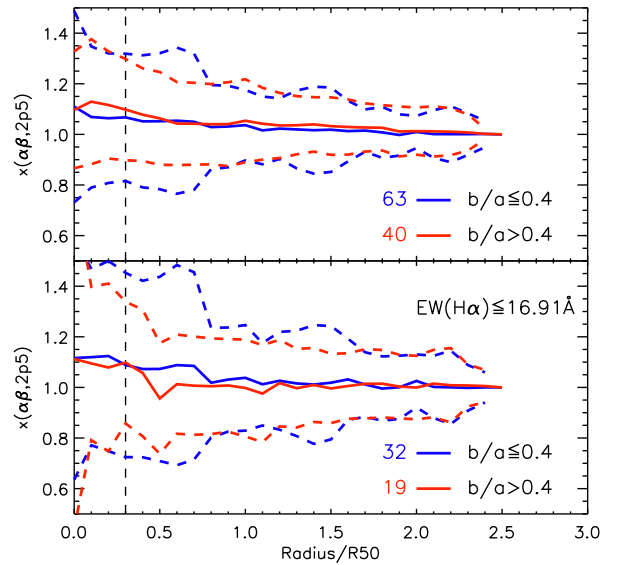}
   \caption{{\bf Top:} Growth curve of $f$(H$\alpha$)/$f$(H$\beta$) normalized to $f$(H$\alpha$)/$f$(H$\beta$) at $2.5 R_{50}$, $x(\alpha\beta,2p5)$, as a function of the radius of the aperture. Lines and colors as in figure~\ref{ha}. 
For one of the galaxies the H$\beta$ emission line was not detected in all the apertures and it was not included.
{\bf Bottom:} Same as top panel restricted to galaxies with $EW$(H$\alpha$) $\leq 16.91$\AA.}
              \label{hahb}%
    \end{figure}

\begin{figure}[!t]
   \centering
\includegraphics[width=9cm]{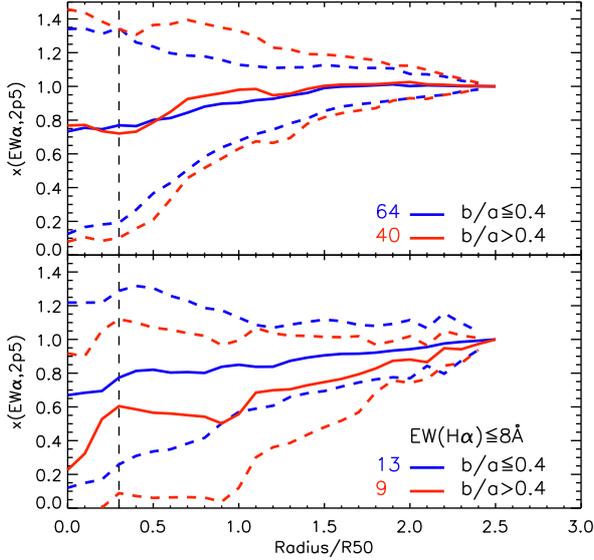}
   \caption{{\bf Top:} Growth curve of $EW$(H$\alpha$) normalized to $EW$(H$\alpha$) at $2.5 R_{50}$, $x(EW\alpha,2p5)$, as a function of the radius. {\bf Bottom:} Same as top panel restricted to the galaxies with $EW$(H$\alpha$)$\leq 8$\AA\ at $2.5 R_{50}$.}
              \label{ewha}%
    \end{figure}

We illustrate our results with the aperture corrections for three observables: $f$(H$\alpha$), the $f$(H$\alpha$)/$f$(H$\beta$) ratio, and the $EW$(H$\alpha$), shown in figures~\ref{ha} to~\ref{ewha}.
The figures show the median profiles of the growth curves corresponding to these three quantities normalized to the value measured within an external isophote containing most of the optical light of the galaxy.
We normalized the profiles to the value measured at $2.5R_{50}$ since this is the typical value covered by the PPAK field-of-view for the CALIFA sample.
This aperture, $2.5 R_{50}$, encloses on average $\simeq 90$\% of the total H$\alpha$ flux of spiral galaxies\footnote{Estimated by adding up the contribution to the H$\alpha$ emission flux of all the spaxels of the PPAK aperture for each galaxy of the sample. This must be taken as a rough estimation since the largest galaxies in the sample show H$\alpha$ emission beyond the PPAK aperture.}.

Figure~\ref{ha} shows the median $f$(H$\alpha$) growth curve normalized to $f$(H$\alpha$) at $2.5R_{50}$, $x(\alpha,2p5)$.
We have split our sample according to the axis ratio in order to take into account the effect that the inclination of the galaxy may have on the values measured through circular apertures.
For this, we use the isophotal axis in the $r'$ band $isoA\_r$ and $isoB\_r$ from SDSS-DR7. 
This choice gives average values of $\left<isoB\_r/isoA\_r\right> = 0.27$ and 0.53 respectively for each subsample.
We have verified that other choices like the semi-minor to semi-major axis from elliptical growth curve and light momentum analysis yield basically the same average inclination distributions as our selection.
As it can be seen, the average curves for both ranges of inclination are quite similar within the $1\sigma$ uncertainties.
The low thickness of the ionized gas disks, as estimated from the low velocity dispersion of the H{\sc ii} regions ($\approx 10~km~s^{-1}$, Rela\~{n}o et al. 2005), suggests that no important differences in the H$\alpha$ growth curves of face-on and edge-on galaxies should be expected. However, a circular aperture covers different parts of the disk for face-on and edge-on galaxies.
Thus, the similarity found for both curves could be partly due to the fact that the average inclination values of our subsamples are far from the extreme face-on and edge-on cases and close enough from each other, although we have verified that even the curves restricted to galaxies with the most extreme values in our sample are not significantly different from each other.
A physically motivated study of the H$\alpha$ spatial distribution for face-on and edge-on galaxies is out of the scope of this work and will be part of ongoing work devoted to the SFR in CALIFA galaxies.
Similarly, aperture corrections as a function of Hubble type require a larger sample and will be part of future papers on this topic.
We list in table~\ref{ha_error} the median growth curve and the $1\sigma$ uncertainties of $f$(H$\alpha$) as a function of the aperture radius.
Given that galaxies with low and high inclination show very similar results, these values correspond to our complete working sample. 

\begin{table}
\caption{
$x(\alpha,2p5)$ and $x(\alpha\beta,2p5)$ and their $1\sigma$ deviations as a function of the size of the circular aperture normalized to $R_{50}$.
}             
\label{ha_error}      
\centering                          
\begin{tabular}{ccccc}        
\hline\hline                 
$r/R_{50}$ & $x(\alpha,2p5)$ & ($\sigma$) & $x(\alpha\beta,2p5)$ & ($\sigma$) \\    
\hline                        
0.3 & 0.091 & (0.074) & 1.078 & (0.221) \\
0.5 & 0.194 & (0.119) & 1.052 & (0.240) \\
0.7 & 0.301 & (0.156) & 1.042 & (0.234) \\
0.9 & 0.419 & (0.158) & 1.034 & (0.163) \\
1.1 & 0.530 & (0.146) & 1.028 & (0.136) \\
1.3 & 0.629 & (0.130) & 1.021 & (0.145) \\
1.5 & 0.718 & (0.117) & 1.026 & (0.154) \\
1.7 & 0.798 & (0.101) & 1.020 & (0.093) \\
1.9 & 0.854 & (0.079) & 1.009 & (0.088) \\
2.1 & 0.917 & (0.070) & 1.009 & (0.129) \\
2.3 & 0.961 & (0.039) & 1.003 & (0.086) \\
2.5 & 1.000 & -- & 1.000 & -- \\
\hline                                   
\end{tabular}
\end{table}

An issue that deserves discussion is the effect of the FWHM of the CALIFA psf ($\sim 3.6$'') on the growth curves of the emission lines.
It was demonstrated by Trujillo et al. (2001) that the parameters of a S\'{e}rsic profile are affected by the seeing. For exponential disks, which correspond to a $n=1$ profile, the central intensity and effective radius are recovered within 20\% of their true value disregarding the ellipticity as far as $FWHM/r_{eff} \leq 0.3$. This result prevents the use of aperture corrections below this limit since they could be severely affected by seeing. Given that the average value of $R_{50}$ for our sample is $\approx 10''$, the range of validity of our aperture correction is valid for radii $r/R_{50} \gtrsim 0.3$.
A further limitation is the fact that the observed size of the galaxies decreases as we move to higher redshifts, and this imposes both a lower and an upper limit to the range of magnitudes for which our correction can be applied. Based on average measurements of $R_{50}$ for spiral galaxies from SDSS, we have derived the validity ranges of the aperture corrections at different redshifts. 
In table~\ref{aper_sdss}, we show these ranges for two representative projects providing spectra with different aperture sizes: SDSS (single-fiber, $\varnothing = 3''$) and SAMI
(multi-fiber bundle, $\varnothing = 15''$, Bryant et al 2011).
In both cases the lower and upper magnitude limits correspond to the magnitudes at which $r_{aper}/R_{50} \geq 0.3$ and $r_{aper}/R_{50} \leq 2.5$ respectively\footnote{There is a further restriction to the upper magnitude limit due to the effect of the psf on the determination of $R_{50}$.
This upper limit corresponds to the magnitude at which $FWHM \approx R_{50}$ and it must apply when it is more restrictive than the one given in table~\ref{aper_sdss} for each redshift.
In the particular case of SDSS images, $R_{50}$ are not corrected for the psf such that this limit can become relevant at high redshifts.
A discussion on the limits of reliability of $R_{50}$ at different redshifts can be found in Gonz\'{a}lez-P\'{e}rez et al. (2011).}
based on their average sizes at the different redshift proved with a sample of SDSS-DR7 galaxies.
The applicability of the aperture corrections is meaningful for individual galaxies within the limits listed in table~\ref{aper_sdss},
thus average corrections to large sample of galaxies in order to correct quantities like the star formation density from observed H$\alpha$ luminosities are not recommended.

\begin{table}[!t]
\caption{
Redshift and absolute magnitude ranges at which the aperture corrections listed in table~\ref{ha_error} are valid for SDSS and SAMI data.
}             
\label{aper_sdss}      
\centering                          
\begin{tabular}{ccc}        
\hline\hline                 
Redshift & \multicolumn{2}{c}{Absolute magnitude range} \\    
         & SDSS & SAMI \\
\hline                        
0.02 & $-18 \lesssim M_{r} $                    & $-22.5 \lesssim M_{r} \lesssim -16.5$ \\
0.04 & $-20.5 \lesssim M_{r}$                   & $-22 \lesssim M_{r} \lesssim -18.5$ \\
0.06 & $-21.5 \lesssim M_{r} \lesssim -16 $     & $M_{r} \lesssim -20$ \\
0.08 & $-22 \lesssim M_{r} \lesssim -16.5 $     & $M_{r} \lesssim -21$ \\
0.10 & $-22.5 \lesssim M_{r} \lesssim -17 $     & $M_{r} \lesssim -21.5$ \\
0.14 & $-22.5 \lesssim M_{r} \lesssim -18 $     & \\
0.18 & $-22.5 \lesssim M_{r} \lesssim -19 $     & \\
\hline                                   
\end{tabular}
\end{table}

Figure~\ref{hahb} (top panel) shows the median $f$(H$\alpha$)/$f$(H$\beta$) ratio growth curve normalized to $2.5 R_{50}$, $x(\alpha\beta,2p5)$, again split for low and high inclination galaxies.
Both curves show a smooth decline from a central value of $\simeq 1.1$ down to $1.0$ at the normalization radius, and no significant difference with the inclination of the galaxies.
The median growth curve and the $1\sigma$ uncertainty for the whole sample are listed in table~\ref{ha_error}.
The very smooth decline of $x(\alpha\beta,2p5)$ supports the earlier result that the dust attenuation shows little dependence with the radial distance (S\'{a}nchez et al. 2012b). 
We examined the effect of non-negligible residual H$\beta$ absorption after subtraction of the underlying stellar continuum by restricting the analysis to galaxies with $EW$(H$\alpha$) $\leq 16.9$\AA\ (see lower panel of Fig. ~\ref{hahb}).
It can be seen that this growth curve is consistent within the uncertainties with the corresponding to the whole sample, what suggests that our H$\beta$ fluxes, and thus the $f$(H$\alpha$)/$f$(H$\beta$) ratio, are on average free from this effect. 

Figure~\ref{ewha} (top panel) shows the median $EW$(H$\alpha$) growth curve, $x(EW\alpha,2p5)$, 
which as in previous cases is similar for galaxies with low and high inclination.
This curve grows with radius, ranging from a central value of $\simeq 0.7$ to $1$ at the normalization radius.
It is remarkable the large dispersion of this quantity measured at small apertures ($\sigma > 0.3$ at $r/R_{50} \leq 1.3$).
The observed dispersion at small radii is due to the fact that $EW$(H$\alpha$) is an intensive quantity (contrarily to H$\alpha$ or H$\beta$ fluxes which are extensive quantities), 
and thus very low or high values at small radii can be measured due to the presence of features like circumnuclear rings of star formation or nuclear starbursts respectively.
This large dispersion at small radii prevents the use of a reliable aperture correction for the $EW$(H$\alpha$). 
The main implication of the large dispersion at small radii is the possibility that star-forming galaxies observed through a small aperture are misclassified as quiescent if the classification is based only in an extrapolation of the central value of $EW$(H$\alpha$).
To illustrate this we show in the bottom panel of figure~\ref{ewha} $x(EW\alpha,2p5)$ as a function of radius restricted to galaxies with $EW$(H$\alpha$) $\leq 8$\AA.
The curve of the low-inclination galaxies is clearly
below that of the high-inclination ones, which follows a trend similar to that of the whole sample.
As an example, according to this plot a face-on spiral galaxy with (total) $EW$(H$\alpha$) $= 6$\AA\ and observed through an aperture covering less than $R_{50}$ will have $\simeq 50$\% chance of showing (observed) $EW$(H$\alpha$) $\leq 3$\AA, and thus being classified as quiescent depending on the adopted limit (previous works separate active and quiescent galaxies at $EW$(H$\alpha$) $\simeq 2-4$\AA, see Cid Fernandes et al. 2010). 
This effect is less dramatic for edge-on galaxies.

We conclude that the median $f$(H$\alpha$) growth curve of spiral galaxies is well defined and shows a low dispersion for a range of aperture sizes between $0.3 R_{50}$ and $2.5 R_{50}$.
The $f$(H$\alpha$)/$f$(H$\beta$) ratio growth curve presents a smooth behavior suggesting low dependence of the dust attenuation with radius.  
Median aperture corrections for this quantity are also meaningful.
The observed large dispersion around the median $EW$(H$\alpha$) growth curve prevents the derivation of a reliable aperture correction especially at small radii.
This large dispersion warns about the use of the $EW$(H$\alpha$) as the criterion to classify star-forming and quiescent galaxies for which the size of the aperture is $\lesssim R_{50}$ since it might result in misclassification of low $EW$(H$\alpha$) galaxies.
We end up by adding that this paper presents only the tip of the iceberg.  
Aperture effects are widely mentioned but rarely understood. Their influence can seriously affect scientific results 
through systematic error (Bland-Hawthorn et al 2011).
Forthcoming papers will be devoted to providing a complete set of diagnostics involving other indicators of SFR and/or abundances, and to applying these results to existing datasets.

\begin{acknowledgements}
This study makes use of the data provided by the 
Calar Alto Legacy Integral Field Area (CALIFA) survey (http://califa.caha.es/).

The
CALIFA collaboration would like to thank the IAA-CSIC and MPIA-MPG as
major partners of the observatory, and CAHA itself, for the unique
access to telescope time and support in manpower and infrastructures. 

The CALIFA collaboration thanks also 
the CAHA staff for the dedication
to this project.

Based on observations collected at the Centro Astron\'omico Hispano
Alem\'an (CAHA) at Calar Alto, operated jointly by the
Max-Planck-Institut f\"ur Astronomie and the Instituto de Astrof\'\i sica de
Andaluc\'{\i}a (CSIC).

We thank the {\it Viabilidad, Dise\~no, Acceso y Mejora } funding program
ICTS-2009-10, for supporting the initial developement of this project. 
JI-P, JVM, CK and AMI thank to the Spanish PNAYA projects “Estallidos” AYA2010-21887-C04-01.
RAM was funded by the spanish programme of International Campus of Excellence Moncloa (CEI).
JF-B acknowledges support from the Ram\'{o}n y Cajal programme by the Spanish Ministry of Economy and Competitiveness (MINECO). This work has been supported by the Programa Nacional de Astronom\'{\i}a y Astrof\'{\i}sica of MINECO, under grants AYA2010- 21322-C03-01 and AYA2010-21322-C03-02. 
LG and VS acknowledge financial support from Funda\c{c}\~{a}o para a Ci\^{e}ncia e a Tecnologia (FCT) under program Ci\^{e}ncia 2008 and the research grant PTDC/CTE-AST/112582/2009.
The Dark Cosmology Centre is funded by
the Danish National Research Foundation. AG acknowledges funding from the European Union Seventh Framework Programme (FP7/2007-2013) under grant agreement n. 267251.

This research has made use of the NASA/IPAC Extragalactic Database (NED) which is operated by the Jet Propulsion Laboratory, California Institute of Technology, under contract with the National Aeronautics and Space Administration. 
\end{acknowledgements}

\end{document}